\documentstyle[aps,prb,psfig,floats]{revtex}
\begin{document}
\twocolumn
\draft
\wideabs{

\title{Penetration of Josephson vortices and  measurement of the
$c$-axis penetration depth in $\bf Bi_{2}Sr_{2}CaCu_{2}O_{8+\delta}$:
Interplay of Josephson coupling, surface barrier and defects}

\author{H.\ Enriquez, N.\ Bontemps}
\address{Laboratoire de Physique de la Mati\`ere Condens\'ee, Ecole Normale
Sup\'erieure, 24 rue Lhomond, 75231 Paris Cedex 05, France}

\author{A.\ A.\ Zhukov, D.\ V.\ Shovkun, M.\ R.\ Trunin}
\address{Institute of Solid State Physics RAS, 142432 Chernogolovka, Moscow
district, Russia}

\author{A.\ Buzdin, M.\ Daumens}
\address{Laboratoire de Physique Th\'eorique, Universit\'e Bordeaux I,
33405 Talence Cedex, France}

\author{T.\ Tamegai}
\address{Department of Applied Physics, The University of Tokyo,
Hongo, Bunkyo-ku, Tokyo 113-8656, Japan}

\maketitle

\begin{abstract}
The first penetration field  $H_{J}(T)$ of Josephson vortices is
measured through the onset of microwave absorption in the locked state,
in slightly overdoped  $\rm{ Bi_{2}Sr_{2}CaCu_{2}O_{8+\delta}}$ single
crystals ($T_{c}\approx 84~K$). The magnitude of $H_{J}(T)$ is too large
to be accounted for by the first thermodynamic critical field $H_{c1}(T)$.
We discuss the possibility of a Bean-Livingston barrier, also supported by
irreversible behavior upon flux exit, and the role of defects, which
relates $H_{J}(T)$ to the $c$-axis penetration depth $\lambda_{c}(T)$.
The temperature dependence of the latter, determined by a cavity
perturbation technique and a theoretical estimate of the defect-limited
penetration field are used  to deduce from $H_{J}(T)$ the absolute value
of $\lambda_{c}(0)=(35 \pm 15)~\mu m$.
\end{abstract}

\pacs{PACS numbers: 47.32Cc, 74.72h, 78.70Gq}
}

\section{Introduction}

The phenomenological Lawrence-Doniach model is generally used to
describe a stack of Josephson-coupled superconducting layers
\cite{ld,clem1,clem2}. This interlayer Josephson tunneling has been
established experimentally by dc or ac Josephson effect experiments
in numerous high-$T_c$ superconductors \cite{kleiner} and is proposed
as a candidate mechanism for superconductivity \cite{anderson}. Such
discrete layered structures have some striking incidence on many
properties:

i) Josephson vortices appear for field parallel to the layers, and in
case their penetration  in this quasi-2D system is impeded by a surface
barrier \cite{BeLi}, the penetration field, henceforward noted
$H_e^{2D}(T)$, is simply inversely proportional to the $c$-axis
penetration depth $\lambda_{c}(T)$ \cite{buzdin1}, unlike isotropic
superconductors (where it is of the order of the thermodynamic critical
field). The occurence of such a barrier was discussed mostly in the
framework of low-field magnetization measurements performed in fields
parallel to the layers in  NdCeCuO \cite{zuo1}, Tl-2201 \cite{zuo2} and
Bi-2212 \cite{niderost}. The quantitative estimates of $\lambda_{c}(T)$
deduced from these data were however disputed \cite{hussey}.

ii) $\lambda_c(T)$ is directly related to the critical current density
between the layers, $J_0(T)$, and is inversely proportional to
the Josephson plasma frequency  $\omega_{ps}$ \cite{bulaevskii,koshelev}.
Both quantities ought to be discussed within the same theoretical background
\cite{soninc,dewilde}. Direct determination of the plasma frequency was
performed through infrared reflectivity measurements in
$\rm La_{2-x}Sr_{x}CuO_{4}$ \cite{tamasaku},
$\rm YBa_{2}Cu_{3}O_{6+\delta}$ \cite{homes}, Tl-2212 or Tl-2201
\cite{Dul} and from microwave
absorption measurements in underdoped Bi-2212 and Bi-2201 \cite{Gaif}.
A large body of literature reported  a sharp microwave absorption line in
presence of a static field applied parallel to the $c$-axis
\cite{matsuda1,tamegai}.  This absorption line was assigned to Josephson
plasma resonance, whose frequency is modified by the field-dependent
interlayer phase coherence \cite{bulaevskii,matsuda2}. However, this
 interpretation is still controversial \cite{sonin}. Although the geometry
of the experiments reported here is different (the external field is
parallel to the ab planes), the specific field dependence of
$\lambda_c$ or $\omega_{ps}$ may be involved, as discussed elsewhere
\cite{matsup,soninc,dewilde}.

Therefore, an independent measurement of the absolute value of $\lambda_c$
(in zero applied field) is of interest.
To date, all the above mentioned properties have been studied
separately.  It is the aim of this paper to bring together two different
microwave measurements in order to obtain the absolute value of
$\lambda_c(T)$:  i) the first penetration field of Josephson vortices is
measured and shown to be related to  $\lambda_c(T)$, (ii) a cavity
perturbation technique \cite{Tru1} is used to determine the temperature
variation of $\Delta \lambda_{ab}(T)$ and $\Delta \lambda_c(T)$
\cite{trunin}.

In the present paper, we focus mainly on the investigation of the
penetration of Josephson vortices through surface resistance measurements
at high frequency (10~GHz) in Bi-2212 \cite{bontemps}. The onset of
microwave absorption allows to determine the penetration field
$H_{J}(T)$ of Josephson vortices at different temperatures. The
magnitude of $H_{J}(T)$ and the irreversible behavior of the dissipation
with respect to flux entry and flux exit point at first sight toward a
Bean-Livingston surface barrier impeding the penetration of Josephson
vortices. However, a closer quantitative investigation, which includes
the experimental determination of the variation $\Delta \lambda_{c}(T)$
of the $c$-axis penetration depth, and the theoretical calculation of
the penetration field in the presence of edge or surface defects, leads
us to the conclusion that $H_{J}(T)$ is eventually controlled by such
surface irregularities. Relying on these theoretical estimates, we deduce
from $H_{J}(T)$ the absolute value of the $c$-axis penetration depth.

\section{Experiment}

Microwave dissipation measurements were performed in various (generally
slightly over-doped) BSCCO single crystals shaped into rectangular
platelets of approximate size
$a\times b\times c\simeq 2\times 1\times 0.03$~mm$^3$:
sample A, $T_c=86$~K, has a transition width $\Delta T_{c} \approx 3$~K
(as determined from the range over which the microwave absorption drops
from normal to superconducting state values), sample B with $T_c=84$~K,
$\Delta T_{c}\approx 3$~K, sample C, $T_c=89$~K,
$\Delta T_{c} \approx 3$~K. Two other similar samples (D and E) were used
for checking the onset of microwave dissipation with respect to the surface
quality as discussed below. Finally, the temperature dependence of the
penetration depth was measured in a set of similar samples by a cavity
perturbation technique at 10~GHz and ac-susceptibility at 100~kHz.
The details of these measurements will be discussed elsewhere
\cite{trunin} while here we shall only make use of the temperature
variations of $\Delta \lambda_{ab}(T)$ and $\Delta \lambda_c(T)$.
An example of the temperature dependence of the surface resistance $R_s(T)$
in the $ab$-plane of slightly overdoped ($T_c=84$~K) BSCCO single crystal
is shown in Fig.~1. The extrapolation of this curve to $T=0$ (inset of
Fig.~1) yields estimate $R_{\rm {res}}\approx 110$~$\mu\Omega$, which is,
to the best of our knowledge, the lowest value ever obtained in BSCCO
single crystals at 10~GHz. The inset of Fig.~1 displays also the linear
change with temperature ($T<50$~K) of
$\Delta\lambda_{ab}(T)=\lambda_{ab}(T)-\lambda_{ab}(5K)$. This linear
variation at low $T$ was previously observed in optimally
doped \cite{Jac1,Lee,Shib2} and slightly overdoped \cite{Shib2} BSCCO
single crystals. Both of the above mentioned parameters of the sample
suggest that the quality of the cuprate planes is fairly high. We note
that the slope of $\Delta\lambda_{ab}(T)$ in the inset of Fig.~1 is
fairly large ($\approx 25$~\AA/K). It could be a consequence of doping
\cite{Shib2,Coch} with respect to optimally doped crystals
\cite{Jac1,Lee}.

\begin{figure}[h]
\centerline{\psfig{figure=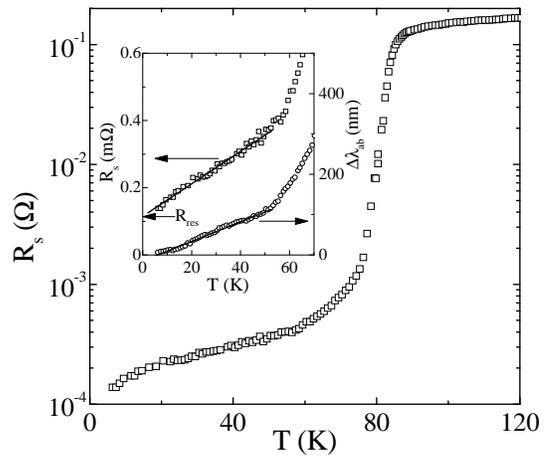,height=6cm,width=7cm,clip=,angle=0.}}
\caption{Surface resistance $R_{s}(T)$ in the $ab$-plane of slightly
overdoped BSCCO single crystal. The inset shows the low temperature
behavior of  $R_{s}(T)$ and of the penetration depth
$\Delta \lambda_{ab}(T)$.}
\label{Rs}
\end{figure}

All samples from different
batches exhibit very similar properties as far as the magnitude and
temperature dependence of the field penetration is concerned, so that
experimental results are only displayed for sample A. $\Delta
\lambda_{c}(T)$ differs among samples with different $T_c$. We shall
only make use of the data on the samples with the same $T_c$ ($\pm 1$~K).

The experimental set-up was described elsewhere \cite{enriquez1}.
It is used to measure the microwave losses as a function
of the applied magnetic field (0-100~Oe) and temperature (50-90~K,
measurements at temperatures lower than 45~K are hindered by the
increasing noise  of the set-up).

The microwave field $\bf {h_1}$ lies within
the $ab$-plane, so that the induced microwave currents flow both within
the $ab$-plane and along the crystallographic $c$-axis. The static
magnetic field $\bf H$ is applied in the $ab$-plane perpendicular to
the microwave field.
A computer-controlled goniometer
allows to select its orientation $\theta$ with respect to the $ab$-plane.
To locate the $\theta=0$ position, we take advantage of the lock-in
transition evidenced earlier \cite{enriquez1}. The set-up measures the
variation of the power dissipated in the cavity as the magnetic field is
swept at fixed temperature, hence yields the field induced imaginary part
$\chi ''(H)$ of the macroscopic susceptibility \cite{enriquez2} (as long as
the dissipation is ohmic, the so-called linear regime). This latter point
has been checked for all the data shown henceforward.

\section{Results}

Figure 2 displays the change of dissipation $\chi''(H)-\chi''(0)$,
starting from zero field (within $\pm 0.1~ \rm Oe$) measured in sample A
for various orientations of the applied field close to the $ab$-plane:
$0^{\circ} \leq \theta \leq 3^{\circ}$ (only the $0^{\circ}$ and
$ 2^{\circ}$ are displayed in Fig.~2) and in a low-field range:
$0~ \leq H \leq 25~\rm Oe$, at three typical temperatures
($T=\rm 78~K, 65~K, 50~K$).

\begin{figure}[h]
\centerline{\psfig{figure=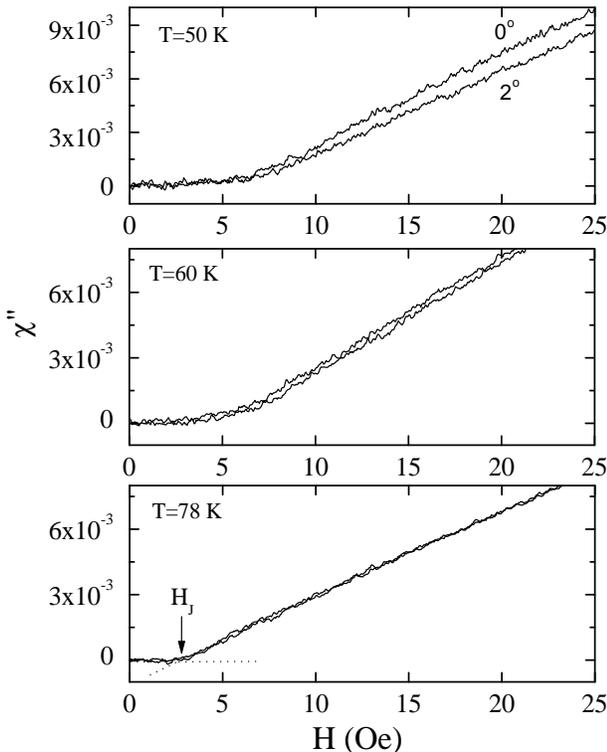,height=10cm,width=8cm,clip=,angle=0.}}
\caption{Dissipation as a function of the applied field at 3 temperatures,
for 2 orientations ($0^{\circ}$ and $ 2^{\circ}$) of the applied
field with respect to the $ab$-plane. The onset of dissipation, indicated
by the arrow, occurs at the penetration field $H_{J}(T)$.
\label{chi}}
\end{figure}

After each field sweep, the sample was warmed through
$T_c$ and then cooled again in zero field, in order to avoid any
possible vortex pinning when studying the penetration starting from zero
field. The dissipation of Josephson vortices is characterized by the fact
that it does not depend on the angle (Fig.~2), as long as these vortices
remain locked. According to our previous study, the dissipation regime
displayed in Fig.~2 comes only from locked Josephson vortices
\cite{enriquez1} and holds up to $\sim 30$~Oe.

As the field increases, an onset in the dissipation occurs at a
temperature-dependent field $H_{J}(T)$ (Fig.~2), which we associate to
Josephson vortices entering the sample. Interestingly, above
$H_{J}(T)$, the microwave absorption behaves linearly with field, with a
very good accuracy, from typically 10~Oe up to 25~Oe. This appears
consistent with a flux-flow mechanism driven by $c$-axis currents, where
the flux-flow resistivity is linear with applied field.
We therefore identify $H_{J}(T)$ to the first penetration field of
Josephson vortices. In this work, unlike in Ref.~\cite{bontemps}, we have
averaged the data over the field orientations from $0^{\circ}$ to
$3^{\circ}$, in an attempt to improve the accuracy of the determination.
As in Ref.~\cite{bontemps}, we choose to define
$H_{J}(T)$ as the field value where the microwave absorption exceeds
the experimental accuracy $(2\cdot 10^{-4})$. The field thus determined
is plotted in Fig.~3. The error bars take into account both the noise and
the estimated drift of the signal with time \cite{note1}.

\begin{figure}[h]
\centerline{\psfig{figure=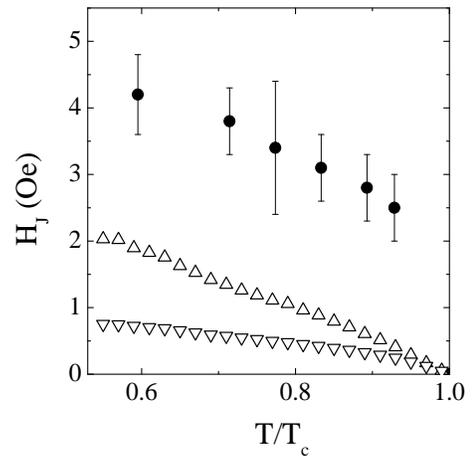,height=6cm,width=6cm,clip=,angle=0.}}
\caption{Plot of $H_{J}(T)$ (full circles). Up (down) triangles display
estimates of $H_{c1}(T)$ using $\lambda_{c}(0)=10~\mu$m ($40~\mu$m). The
temperature variations $\Delta \lambda_{ab}(T)$ and $\Delta \lambda_{c}(T)$
are taken from our present work (Figs.~1,5).
\label{HJ}}
\end{figure}

\section{Discussion}

Whether  $H_{J}(T)$  may be identified to the thermodynamic lower
critical field $H_{c1}$ was previously discussed \cite{bontemps,enriquez1}.
For Josephson vortices $H_{c1}$ writes \cite{clem2}:

\begin{equation}
H_{c1}(T)={{\phi_{0} \over {4 \pi \lambda_{ab}(T) \lambda_{c}(T)}}}
[\ln {\lambda_{ab}(T) /d} + 1.12]
\end{equation}

In our early work  \cite{enriquez1}, we had not yet studied the temperature
dependence of $H_{J}(T)$ and we had not observed the irreversible behavior of
the dissipation upon flux entry and flux exit. We had therefore not considered
the possibility of a surface barrier.
However, in order to reconcile the magnitude of $H_{J}(T<<T_{c})$ with the
thermodynamic lower critical field, we were compelled to take the lowest
possible values for $\lambda_{ab}(0)$ and $\lambda_{c}(0)$.

We proposed next in  Ref.~\cite{bontemps} to take more acceptable lower
bounds for $\lambda_{ab}(0)$ and $\lambda_{c}(0)$, together with the
experimentally determined temperature variations in order to obtain an
upper bound for $H_{c1}(T)$.  Here, we take $ 2100$~\AA\ as a lower bound
for $\lambda_{ab}(0)$ \cite{Lee,schilling}, and  $10~\mu$m for
$\lambda_{c}(0)$
\cite{tamegai,farrell,okuda,nakamura,martinez,steinmeyer,iye,tsui,khaykovich}.
We use the temperature dependence
for $\Delta \lambda_{ab}(T)$ (partly shown in the inset of Fig.~1) and
$\Delta \lambda_{c}(T)$ measured in our previous work \cite{trunin}
(see Fig.~5 below).  The corresponding $H_{c1}(T)$ is plotted in
Fig.~3 using the above mentioned values. We have also displayed in
Fig.~3 $H_{c1}(T)$ if taking  $\lambda_{c}(0)=40~\mu \rm m$
\cite{trunin,Jac1}. It is clearly seen that neither the absolute value
(too small compared to the experimental data) nor the temperature
dependence (quasi-linear) agrees with the $H_{J}(T)$ data. Since the
actual penetration field is larger than the thermodynamic $H_{c1}(T)$, it
is therefore quite natural to assume that a Bean-Livingston surface
barrier impedes field penetration, and yields a larger entry field
$H_e^{2D}(T)$.

In  anisotropic superconductors, in the quasi-2D regime, e.g. when the
transverse coherence length $\xi_c$ becomes smaller than the interlayer
distance $d$, $H_e^{2D}(T)$ was shown to be related only to the c-axis
penetration length through \cite{buzdin1} :

\begin {equation}
 H_e^{2D}(T)={\phi_{0} \over {4 \pi \lambda_{c}(T) d}}
\end {equation}

In Bi-2212, the quasi-2D regime holds up to temperatures very close to
$T_c$, so that this last expression for $H_e^{2D}(T)$ is valid in our
measuring temperature range. A surface barrier might thus account for
the observed value of the penetration field. Also, since $H_e^{2D}(T)$
grows as $1/\lambda_{c}$(T) (instead of $1/\lambda_{ab} \lambda_{c}(T)$),
it is expected that the temperature dependence could show a better agreement.
The existence of a surface barrier is further suggested by the hysteretic
behavior of dissipation, shown in Fig.~4, at $T=65$~K (the behavior is
similar at other temperatures). When the field is swept down, vortices do
not exit in a reversible way. However all vortices have left the sample as
can be inferred from the recovery of the same dissipation as in zero
initial field, when the field is back to zero value. When the field is
swept up again, the absorption displays precisely the same behavior
as after the  zero field cooled procedure. Bulk pinning  would
induce flux trapping at zero field, hence some residual dissipation.
Our observations are similar to magnetization measurements where
the irreversibility, assigned to a surface barrier, is characterised
by zero magnetization upon decreasing field. Such a behavior, first
observed in a field parallel to the $c$-axis \cite{kon}, was also
reported for Josephson vortices in Bi-2212 in a field
oriented nearly parallel to the ab plane \cite{nakamura}.
Surface barriers may also lead to time dependent effects \cite{burla}.
Indeed, it was argued from  magnetization data taken at various sweep
rates that the penetration field in the parallel configuration depends on
the field sweep rate, and eventually achieves the thermodynamic first
critical field value for the slowest rates \cite{niderost}. Our sweeping
rate is of the order of 0.1~Oe/s, comparable to the range where the
largest penetration fields were observed \cite{niderost}.
We did not change the sweeping rate hence we cannot confirm this claim.
We point out however that the penetration fields observed in
\cite{niderost} are significantly smaller (roughly a factor of 3) than
ours. Compared to the fastest rate, the decrease of the penetration field
associated to the slowest rate is only 1 Oe. Such small values can obviously
be more easily reconciled with $H_{c1}(T)$ than ours.

\begin{figure}[h]
\centerline{\psfig{figure=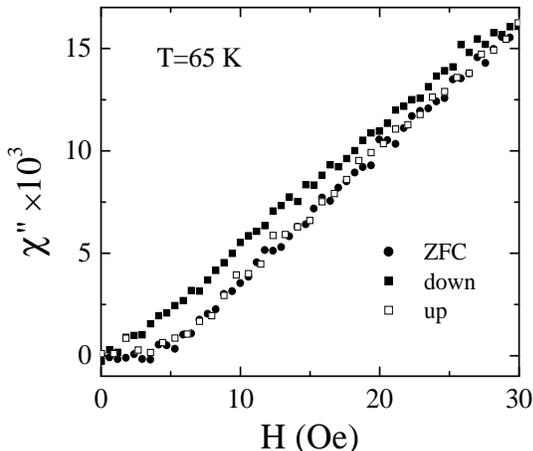,height=6cm,width=7cm,clip=,angle=0.}}
\caption{Plot of $\chi''(H)$ as a function of the direction of the field
sweep. Full circles, full squares and open squares refer to sweeping up
the field starting from a ZFC state, sweeping down the field to zero,
and sweeping up the field again respectively.
\label{hysteresis}}
\end{figure}

It is worth noting that all these remarks do not modify the surface
barrier interpretation: they  only put a time scale for its observation.

Relying on the results described above, we derive from the $H_{J}(T)$ data
an effective penetration depth $\lambda_{J}(T)$ using Eq.~(2). The data
are shown in Fig.~5. We then try to determine $\lambda_{c}(0)$ so as to
fit $\lambda_{J}(T)$ using the measured $\Delta \lambda_{c}(T)$.
We find that both sets of data, namely $\lambda_{J}(T)$ and
$\Delta \lambda_{c}(T)$ cannot be reconciled for any value we may assume
for $\lambda_{c}(0)$. Therefore, the interpretation cannot be so simple.

\begin{figure}[h]
\centerline{\psfig{figure=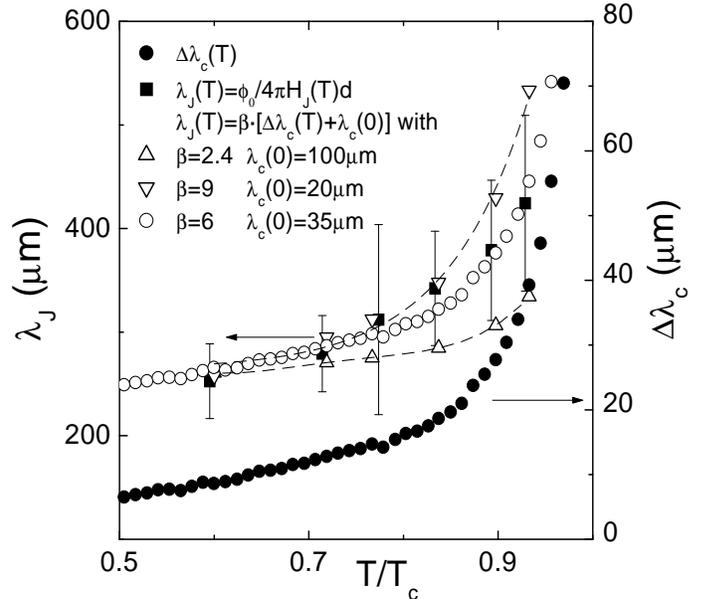,height=8cm,width=9cm,clip=,angle=0.}}
\caption{Plot of the temperature variation $\Delta \lambda_{c}(T)$
(full circles, right scale) and of the effective length $\lambda_J(T)$
which is associated to a surface barrier for the penetration of Josephson
vortices (full squares, left scale), using Eq.~(19) or (21). Open circles
display the best fit using $\lambda_{c}(0)=35~\mu$m and a scaling factor
$\beta=6$. Open symbols show the best fits using $\lambda_c(0)=20~\mu$m
(down triangles) and $\lambda_c(0)=100~\mu$m (up triangles).
\label{lambda}}
\end{figure}

\section{Role of surface irregularities}

\subsection{Experimental checks}

Actually, a surface barrier is only  effective if the surface is smooth on
a typical length scale which is the penetration depth. In our field
geometry, defects either located on the top and bottom, e.g. $ab$-planes,
or the edges may destroy the surface barrier. In the former case, the
relevant length scale is $\lambda_{ab}(T)$, in the latter case,
$\lambda_{c}(T)$.

In order to distinguish between these two possibilities, we have carried
out several checks.  The samples discussed in this paper were first
measured without any special preparation except for their initial shaping
in platelet and cleaving in order to work on a well defined single crystal
and mirror-like surfaces. We noticed that cleaved surfaces often exhibit a
few visible steps and sparse voids.  After the first measurement, sample D
was placed on the stage of an STM and the tip was used in order to cut
four grooves parallel to the small side of the crystal, 4000~\AA\ deep
and  100~$\mu$m apart. Then the sample was measured again ($\bf H$
parallel to the grooves). No significant change in the onset field of the
microwave absorption was observed. In a second step, we took another
sample yielding a similar penetration field, and cleaved it.  We obtained
fresh surfaces with one or two isolated steps which could be seen under a
binocular.  This sample was measured immediately after cleaving, and
again, no significant change was observed. It seems therefore that either
defects within the $ab$-planes do not play any role in order to reduce a
surface barrier or even a single step is immediately effective to destroy
the surface barrier.

One should also consider penetration through the edges. Indeed, the edges
are fairly difficult to control. We did check indirectly, in the surface
impedance and ac-susceptibility experiments, whether they play any role.
In order to measure $\Delta \lambda_{c}(T)$, the rf magnetic
field applied parallel to the plane is also parallel to one edge of the
crystal. If the sample is rotated by $90 ^{\circ}$ along its $c$-axis,
the edges where $c$-axis currents flow are interchanged. It is then clear
that if there exists a large defect, e.g. a slit or groove deep in one
edge and not in the other, this defect changes significantly the  $c$-axis
microwave current pattern in one position and much less in the other.
Therefore, the two configurations should yield a different
$\Delta \lambda_{c}(T)$ result. In one particular sample out of three,
this was indeed the case, suggesting the presence of a defect lying in
one edge and showing that the measured $\Delta \lambda_{c}(T)$ cannot be
intrinsic for this particular sample. The data used in this paper and
shown in Fig.~5 are not biased by such large edge defects, e.g.
the  $\Delta \lambda_{c}(T)$ data displayed in Fig.~5 are unchanged
within the accuracy of the measurement upon this rotation.

We have now to examine quantitatively to which extent defects located
on the top or bottom surfaces, or in the edges
alter  the penetration field.

\subsection{Theoretical calculations: formalism}

As usual the entrance field is deduced from the balance between the vortex
attraction to the surface and the pushing force exerted by the screening
current at the minimum distance $\xi $ (the vortex core size) \cite
{BeLi,DeGen}. The presence of the surface irreqularities can strongly
influence the screening current distribution. In particular, near a
scratch the current density can be many times larger than near the flat
surface. This may substantially increase the force pushing vortices
inside the superconductor and then decrease the surface barrier and
the entrance field. The vortex attraction to the surface does not
change essentially near a scratch, as it has been demonstated in
Ref.~\cite{BuzDau}. The force of attraction can decrease by at most a
factor of two near the defect. Then, the main change of the entrance
field is essentially related with the increase of the screening current
density.

We consider the case where the scratch is in the form of a groove on the
superconductor surface, and the magnetic field is parallel to it. Let the
$z$-axis be perpendicular to the superconductor surface. The magnetic
field is parallel to this surface along the $y$-axis, and we choose the
axis of the groove on the same direction. The depth of the scratch is
denoted as $b$ and its width $2a$ (see Fig.~6). For convenience, the
semi-axis $z>0,$ is chosen inside the superconducting material, so in
Fig.~6 the scratch is presented on the bottom surface of the
superconductor. Both $a$ and $b$ are supposed to be much smaller than
$\lambda$, the London penetration depth, so screening can be ignored and
the two-dimensional London equation reduces to Poisson's equation. Then
the lines of current correspond to the equipotentials, and a dielectric
defect in a superconductor corresponds to a metallic embedding in
electrostatics \cite{BD2}. This analogy reduces our problem to the
calculation of the electric field distribution near a metallic electrode
having the special form (Fig.~6) while the field becomes uniform for
$z\rightarrow \infty$. As is known from electrostatics (see,
e.g., Ref.~\cite{Dur}), the solution is provided by a conformal
transformation of the $w$-plane, corresponding to a flat surface, to the
$\zeta$-plane, the plane of an orthogonal cut of the scratch. In the
$w$-plane the attraction energy of the plane on a vortex located at the point
$w$ can be easily computed, for example by the image method \cite{DeGen}

\begin{figure}[h]
\centerline{\psfig{figure=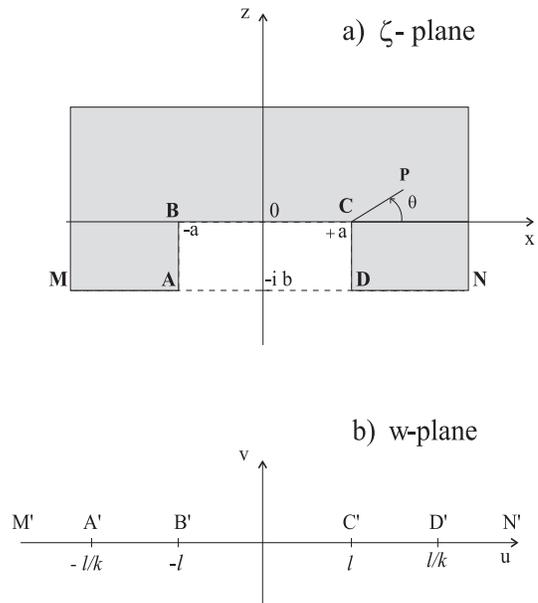,height=8cm,width=7cm,clip=,angle=0.}}
\caption{
a) The defect MABCDN in the form of the groove at the bottom
surface of superconductor. The plane of the figure corresponds to the plane
$\zeta =x+iz$. b) The plane $w=u+iv$ where the straight line
M$'$A$'$B$'$C$'$D$'$N$'$ is mapped to the surface line MABCDN.
\label{calcul}}
\end{figure}

\begin{equation}
E_{att}(w)=-\left( \frac{\phi _{0}}{4\pi \lambda }
\right) ^{2}\ln \frac{\lambda }{w-\overline{w}}\,.
\label{tc1}
\end{equation}

Besides, a uniform current density $j(w)=j_{{0}}$ in the $w$-plane,
can be deduced from the simple complex potential $\psi (w)=j_{0}\,w$.
Then in the $\zeta$-plane, the complex current density $j(\zeta)$ can
be obtainded from the complex potential $\psi (\zeta )=j_{0}\,w(\zeta)$ by

\begin{equation}
j(\zeta )=\frac{d\psi }{d\zeta }=j_{0}\frac{dw(\zeta )}{d\zeta },
\label{tc2}
\end{equation}
where $j_{0}$ is the current density far away
from the defect, i.e., the screening current near the surface
$j_{0}=cH/4\pi \lambda$. To calculate both the attraction
energy and the current density in the $\zeta$-plane, we need to inverse
the conformal transformation. In general, this cannot be done
analytically. However, for situations of practical interest, we may use
approximations that allow us to obtain an analytic solution.

\subsection{Isotropic case}

In the Appendix, we have demonstrated that according the values of
$a,b,\zeta$, there are three different regimes:

(i) $a\ll b\sim \left| \zeta \right| $ slit-like defect,

(ii) $\left| \zeta \right| \ll a\ll b\ $groove-like defect,

(iii) $\left| \zeta \right| \ll b\ll a$ step-like defect.

Let us begin by the slit-like defect. In this case, we use equations
(\ref{tc10a}) and (\ref{tc12}) to derive the vortex
attraction energy at the distance $z$ from the slit

\begin{equation}
E_{att}(z)=-\left( \frac{\phi _{0}}{4\pi \lambda }\right) ^{2}\ln
\frac{\lambda }{2\sqrt{2bz}},
\label{tc17}
\end{equation}
and the strength of the attraction force is
\begin{equation}
f_{att}(z)=\frac{1}{2}\left( \frac{\phi _{0}}{4\pi \lambda }
\right) ^{2}\frac{1}{z},
\label{tc18}
\end{equation}
that is half of the force for a plane surface. Similarly the current
density is \begin{equation}
j(z)=\frac{1}{\sqrt{2}}j_{0}\sqrt{\frac{b}{z}}.  \label{tc19}
\end{equation}

These two results were obtained earlier \cite{BuzDau} for the field and
current distribution near the angle 2$\pi $ (the cut at the superconductor
surface).

The balance between the vortex attraction to the surface and the pushing
force exerced by the screening current at the minimum distance $\xi $ (the
vortex core size) \cite{BeLi,DeGen} gives the entrance field near the defect
\begin{equation}
H_{ed}\simeq H_{e}\left( \frac{\xi }{b}\right) ^{1/2},  \label{tc20}
\end{equation}
where $H_{e}$ is the entrance field for the flat surface
$H_{e}\simeq \phi_0/4\pi \lambda \xi\simeq H_{c}$ (thermodynamic critical
field). The current
concentration effect near the slit essentially reduces the entrance field.
In fact this situation where $z\sim \xi \gg a$ is not realistic for
isotropic superconductors, but it will be useful for the description of
the anisotropic ones.

For the groove-like defect, equations (\ref{tc14}) and (\ref{tc16a}) allow
to derive the physical quantities in the vicinity of the point $C.$ Let $P$
be a point such that
\begin{equation}
\zeta _{P}=a+\rho \,e^{i\theta },\ \rho \ll a.
\end{equation}

The values of $\theta $ are limited by the groove and the core of the vortex
\begin{equation}
-\frac{\pi }{2}+\arcsin (\xi/\rho)\leq \theta \leq \pi -\arcsin (\xi/\rho).
\label{tc22}
\end{equation}

The attraction energy on a vortex at the point $P$ is

\begin{equation} E_{att}(\rho ,\theta )=
-\left(\frac{\phi _{0}}{4\pi \lambda }\right)^{2}
\ln \left[\left(\frac{2}{9b\rho^2 \sqrt{\pi}}\right)^{1/3}
\frac{\lambda}{\sin (\frac{2\theta +\pi }{3})}\right] ,
\label{tc23}
\end{equation}
and the strength of the attraction force reads
\begin{equation}
f_{att}(\rho ,\theta )=\frac{2}{3}\left( \frac{\phi_{0}}
{4\pi \lambda}\right)^{2}\frac{1}{\rho \sin (\frac{2\theta +\pi }{3})}.
\label{tc24}
\end{equation}
Its maximun is obtained for $\theta =0$ or $\pi /2,$ this strengh is reduced
by a factor $4$/$(3\sqrt{3})\simeq 0.77$ by comparing to a flat surface. The
calculation of the current density at the point $P$ gives
\begin{equation}
j(z)=\left(\frac{\sqrt{\pi }}{6}\right)^{1/3}\,
j_{0}\left(\frac{b}{a}\right)^{1/6}
\left(\frac{b}{\rho}\right)^{1/3}e^{-i\theta /3}.
\label{tc25}
\end{equation}

As usual by setting a vortex at a distance $\xi$ near the defect, we obtain
for the entrance field
\begin{equation}
H_{ed}\simeq 2\left( \frac{2}{9\sqrt{\pi }}\right)^{1/3}H_{e}
\left(\frac{a}{b}\right)^{1/6}\left( \frac{\xi }{b}\right)^{1/3},
\label{tc26}
\end{equation}
where $[16/(9\sqrt{\pi})]^{1/3}\simeq 1.$

Finally, for a step $a\gg b,$ by using equations (\ref{tc14}) and (\ref
{tc16b}) we can derive the vortex attraction energy and the current density
at the point $P$ in the vicinity of the point $C.$ The vortex attraction
force is still given by Eq.~(\ref{tc24}) while the current
distribution near the corner becomes

\begin{equation} j(\rho ,\theta
)\simeq \left( \frac{4}{3\pi }\right) ^{1/3}j_{0}\left( \frac{b}{\rho
}\right) ^{1/3}e^{-i\theta /3}.
\label{tc27}
\end{equation}

The corresponding entrance field is
\begin{equation}
H_{ed}\simeq \left(\frac{2\pi }{9}\right) ^{1/3}H_{e}
\left( \frac{\xi }{b}\right) ^{1/3},
\label{tc28}
\end{equation}
where $(2\pi /9)^{1/3}\simeq 0.89.$

\subsection{Anisotropic case}

Now we consider the case of anisotropic superconductors, keeping in mind
layered high-T$_{c}$ materials. As usual, let the $z$-axis (or $c$-axis) be
perpendicular to the superconducting layers. We shall consider two cases :
either the groove is on the bottom surface of the crystal, or it is on the
side surface (edge) of the layered material. In both cases we choose the axis
of the groove parallel to the layers along the $y$-axis, and the magnetic
field in the same direction : $\mathbf{h}$=h(x,z)$\,\mathbf{e}_{y}$. For
such a geometry we may write the London free energy of the anisotropic
superconductor as
\begin{equation}
F=\frac{1}{8\pi }\int \left[\mathrm{h}^{2}+
\lambda_{ab}^{2}\left(\frac{\partial h}{\partial x}\right)^{2}+
\lambda_{c}^{2}\left( \frac{\partial h}{\partial z}\right)^{2}\right] dV\,,
\label{tc29}
\end{equation}
where $\lambda_{c}$ is the London penetration depth when the screening
current is flowing along the $z$-axis ($c$-axis) and $\lambda _{ab}$ when
the current is in $(x,y)$ plane. For a high-T$_{c}$ superconductor, we
have $\lambda_{c}\gg \lambda_{ab}.$

For a very anisotropic superconductor, in the quasi $2D$ regime we have
$\xi_{c}<d,$ where $d$ is the interlayer distance. In such a case, we
need to use $d$ as the size, in the $z$-direction, of the vortex core
\cite{buzdin1} in calculating the entrance field. For the flat surface
the entrance field becomes $H_{e}^{2D}\simeq \phi _{0}/[4\pi d\lambda_c].$

By making a scaling transformation, we introduce a new coordinate :
$X=(\lambda_{ab}/\lambda_c)\,x\ll x$ \cite{BuzDau}. Then the London free
energy (\ref{tc29}) takes the same form as for the isotropic superconductor
with the London penetration depth $\lambda_{ab}$ and we can use the
results of the previous section.

Let us consider the case when the groove is on the bottom surface of the
crystal. Under the scaling transformation, the width of the groove changes
\begin{equation}
a\rightarrow a^{\prime }=\frac{\lambda _{ab}}{\lambda _{c}}\,a\ll a.
\label{tc30}
\end{equation}

Then, the entrance field will be given by the corresponding formulas for
the isotropic case with the replacement of $a$ by $a^{\prime }$ and
$\xi $ by $\xi _{c}\,,$ the correlation length along the $z$-axis
(or by $d$ when $\xi_{c}<d)$.

For $d\sim b\gg a^{\prime }$, the groove may be considered simply as a thin
cut at the surface and by using Eq.~(\ref{tc20}) we derive
\begin{equation}
H_{ed}^{2D}\simeq H_{e}^{2D}\left( \frac{d}{b}\right) ^{1/2}.  \label{tc31}
\end{equation}
Note that due to the large anisotropy of some high-T$_{c}$ materials, this
situation could be realized in practice despite the fact that $d$ is of the
order of only 10 \AA\ .

In the opposite case$\,d\ll a^{\prime }\ll b$, Eq.~(\ref{tc26}) gives the
entrance field near the groove
\begin{equation}
H_{ed}^{2D}\simeq H_{e}^{2D}\left( \frac{d}{b}\right) ^{1/3}\left( \frac{%
a\lambda _{ab}}{b\lambda _{c}}\right) ^{1/6}.  \label{tc32}
\end{equation}
Near the step $a^{\prime }\gg b\gg d,$ the entrance field becomes
\begin{equation}
H_{ed}^{2D}\simeq H_{e}^{2D}\left( \frac{d}{b}\right) ^{1/3}.  \label{tc33}
\end{equation}

Finally, if the groove-like scratch is on the side surface of the layered
material, its effective depth $b^{\prime }$ after the scaling transformation
becomes much smaller
\begin{equation}
b^{\prime }=\frac{\lambda _{ab}}{\lambda _{c}}\,b\ll b.
\label{tc34}
\end{equation}

Then for $b^{\prime }\ll a$ such a scratch has practically no effect on the
vortex entrance. In the opposite case $b^{\prime }\gg a>d,$ by using Eq.~(\ref
{tc26}), we obtain for the entrance field :
\begin{equation}
H_{ed}^{2D}\simeq H_{e}^{2D}\left( \frac{d}{b}\right) ^{1/3}\left( \frac{a}{b%
}\right) ^{1/6}\left( \frac{\lambda _{c}}{\lambda _{ab}}\right) ^{1/2}.
\label{26}
\end{equation}

In fact the lateral defect must be rather deep: $b\gg a\lambda _{c}/\lambda
_{ab}\geq d\lambda _{c}/\lambda _{ab}$ to strongly reduce the entrance
field value. We may deduce that the parallel entrance field depends strongly
on the surface defects in layered superconductors. The current concentration
near the defect edges may greatly reduce the entrance field in comparison to
its theoretical value $H_{e}^{2D}\simeq \phi_{0}/4\pi d\lambda_{c}.$

\subsection{Comparison with experimental data}

We have therefore attempted to fit our data derived from $H_{J}(T)$
with Eqs.~(19), (21) or (23),
using two adjustable parameters: a scaling factor $\beta$ associated with
the defect geometry which equals to $(b/d)^{1/2}$ in Eq.~(19),
$(b/d)^{1/3}$ in Eq.~(21), $(d^2a/b^3)^{1/6}$ in Eq.~(23), and the
absolute value of $\lambda_{c}(0)$. We show the results in Fig.~5, only
for the case described by Eqs.~(19) and (21) (defect in the $ab$-plane),
where we have determined the scaling factor and $\lambda_{c}(0)$ which
allow to adjust $\lambda_{J}(T)$. We find a best fit for
$\lambda_{c}(0)=35$~$\mu$m and a scaling factor $\beta=6$. Assuming a thin
groove, this yields $b \sim 500$~\AA\, which is reasonable. We also show in
Fig.~5 smaller and larger values for $\lambda_{c}(0)$. They allow us to
set the uncertainty about our determination of the penetration depth.

As for Eq.~(23) (edge defect), we have taken $\lambda_{ab}(0)=2600$~\AA\
. We can also account for the data but only in a very restricted,
nevertheless acceptable, range of parameters.
The depth of the edge slit should be of the order of $10~\mu$m, which is
still small with respect to $\lambda_{c}(0)$. The key result in this
latter case is that it yields the same absolute value for
$\lambda_{c}(0)$.

\vspace{0.5cm}

In conclusion, the set of experiments that we have performed suggest very
strongly a surface barrier which impedes field penetration, nevertheless
partially destroyed according to the calculations
developed in the framework of this work. Although we cannot ascertain
which specific defects reduce the efficiency of the surface barrier,
we obtain a fairly good estimate of the $c$-axis penetration depth.

\section{Acknowledgements}

This work was supported by the Centre National de la Recherche Scientifique
- Russian Academy of Sciences cooperation program 4985, and by CREST
and Grant-in-Aid for Scientific Research from the Ministry of
Education, Science, Sports and Culture (Japan).
The work at ISSP was also supported by the Russian
Fund for Basic Research (grant 00-02-17053) and
Scientific Council on Superconductivity (project 96060).

\appendix

\section*{Conformal transformation}

The Schwarz-Christoffel conformal transformation of the $w$ plane $(w=u+iv)$
to the $\zeta $ plane $(\zeta =x+iz)$ which maps the straight line $%
M^{\prime }A^{\prime }B^{\prime }C^{\prime }D^{\prime }N^{\prime }$
(Fig.\thinspace 6b) to the surface line $MABCDN$ (Fig.\thinspace 6a) is
\cite {Dur}
\begin{equation}
\zeta (w)=A\int_{0}^{w/\ell }\sqrt{\frac{1-t^{2}}{1-k^{2}t^{2}}\,dt}\,,
\label{tc3}
\end{equation}
where the two parameters $k$ and $\ell $ are related to the dimensions $a$
and $b$ of the groove, and the constant $A$ is simply
\begin{equation}
A^{-1}=a^{-1}\int_{0}^{1}\sqrt{\frac{1-t^{2}}{1-k^{2}t^{2}}\,dt\,}.
\label{tc4}
\end{equation}

The integrals of the two previous equations can be expressed in terms of
incomplete and complete elliptic integrals
${E}(z,k),\,{F}(z,k),\,{E}(k)\equiv {E}(1,k),$ ${K}(k)\equiv $ ${F}(1,k)$
\cite{Abramo}. For this we define two ${G}$ functions, one incomplete and
one complete as :
\begin{eqnarray}
{G}(z,k)={E}\left( z,k\right)-k^{\prime \,2}{F}\left(z,k\right),
\nonumber \\
{G}(k)={G}(1,k)={E}\left( k\right)-k^{\prime \,2}{K}\left( k\right),
\label{tc5}
\end{eqnarray}
where $k^{\prime }=\sqrt{1-k^{2}}.$ Then the conformal transformation
reads
\begin{equation}
\zeta (w)=a\frac{{G}\left( w/\ell ,k\right) }{{G}\left( k\right) }\,.
\label{tc6}
\end{equation}

The dimensionless parameter $k$ $\in \lbrack 0,1]$ is determined by the
following equation,
\begin{equation}
\frac{a}{b}=\frac{{G}\left( k\right) }{{G}\left( k^{\prime }\right)}\,.
\label{tc7}
\end{equation}

The limits $k\rightarrow 0$ and $k\rightarrow 1$ correspond to
$a/b\rightarrow 0$ and $a/b\rightarrow \infty $ respectively. The last
parameter $\ell,$ the dimension of which is a length, is determined by
requiring that at a large distance from the defect, the two variables
$\zeta$ and $w$ become equal. Using the asymptotic form of ${G}(z,k)$ for
large $z$,
\begin{equation}
\left| z\right| \gg 1/k\Rightarrow {G}(z,k)\simeq kz,
\label{tc7a}
\end{equation}
we get
\begin{equation}
\ell =a\frac{k}{{G}\left( k\right) }\,.
\label{tc8}
\end{equation}

When $k\rightarrow 0$, i.e., $k^{\prime }\rightarrow 1,$ the following
asymptotic forms of the elliptic integrals

\begin{equation}
k\rightarrow 0,{ \ }{G}\left( k\right) \simeq \frac{\pi }{4}k^{2},
{ \ }{G}\left( k^{\prime }\right) \simeq 1,{\ }
\label{tc9}
\end{equation}
may be used to determine the parameters $k$ and $\ell $ in the
limits $a/b\rightarrow 0$ or $a/b\rightarrow \infty $

\begin{eqnarray}
a/b &\rightarrow &0,{ \ }k\simeq \frac{2}{\sqrt{\pi }}\sqrt{\frac{a}{b}}%
\,,\,\,\,\ell \simeq \frac{2}{\sqrt{\pi }}\sqrt{a\,b}\,,
\label{tc10a}
\\
a/b &\rightarrow &\infty ,{ \ }k\simeq 1-\frac{2}{\pi }\frac{b}{a},%
{ \ }\ell \simeq a{ .}
\label{tc10b}
\end{eqnarray}

Firstly, we suppose that the groove is very narrow $a\ll b,$ and we
consider the region where $\zeta \sim w\sim b.$ Then we have
$\left|w\right| \gg \ell $ and the elliptic integrals can be
approximed as
\begin{equation}
k\rightarrow 0,{ }1\ll \left| z\right| \ll 1/k\Rightarrow
{\ }{G}\left( z,k\right) \simeq \frac{i}{2}k^{2}z^{2}.
\label{tc11}
\end{equation}

The conformal transformation becomes simpler and it can be inverted
\begin{equation}
\zeta (w)=\frac{i\,}{2}\frac{w^{2}}{b}\Leftrightarrow w(\zeta )=
e^{i\pi/4}\sqrt{2b}\,w^{1/2}.
\label{tc12}
\end{equation}

Secondly we consider the vicinity of the point $C$ in the $\zeta$-plane
$(\left| \zeta -a\right| \ll a)$ and of $C^{\prime }$ in the $w$-plane
$(\left| w-l\right| \ll \ell )$. In this case we have
$\left| w/\ell-1\right| \ll 1$ and the behaviour of the elliptic
integrals is
\begin{equation}
\left| \eta \right| \ll 1\Rightarrow {G}\left( 1+\eta ,k\right) \simeq
{G}\left( k\right) -i\frac{2\sqrt{2}k^{2}}{3k^{\prime }}\eta ^{3/2}.
\label{tc13}
\end{equation}

As previously the conformal transformation can be easily inverted
\begin{eqnarray}
\zeta (w)-a\simeq \frac{i}{\sqrt{a}}\left( \frac{w-\ell }{\varphi (k)}%
\right) ^{3/2} \Leftrightarrow
\nonumber \\
w(\zeta )-\ell =e^{i\pi /3}\varphi
(k)a^{1/3}(\zeta -a)^{2/3},
\label{tc14}
\end{eqnarray}
where the function $\varphi (k)$ is defined as
\begin{equation}
\varphi (k)=\frac{1}{2}\left[ \frac{9k^{\prime 2}}{k{G}\left( k\right) }%
\right] ^{1/3}.
\label{tc15}
\end{equation}

The asymptotic forms of this function read
\begin{eqnarray}
a/b &\rightarrow &0,{ \ }\varphi (k)\simeq \frac{1}{2}
\left( \frac{9\sqrt{\pi }}{2}\right)^{1/3}\sqrt{\frac{b}{a}}\,,
\label{tc16a} \\
a/b &\rightarrow &\infty ,{ \ }\varphi (k)\simeq
\left( \frac{9}{2\pi}\right)^{1/3}\left( \frac{b}{a}\right)\,.
\label{tc16b}
\end{eqnarray}

Note that in this last limit $a\gg b,$ we retrieve the case of a single step
defect. The second step at the point $C$ is not involved.


\begin{references}
\bibitem{note1} It might appear attractive to take advantage of the
linear, temperature independent, field variation of $\chi''$ in order
to extrapolate the field value for zero dissipation. However,
a closer look at the data show that the shape of $\chi''$ at the onset
is temperature dependent and the onset field is found either
above or below the extrapolated field. This shows that in
the absence of a quantitative understanding of how the dissipation of
Josephson vortices sets in, this linear extrapolation remains speculative.
\end{references}
\end{document}